\documentclass[preprint,amsmath,amssymb,nofootinbib]{revtex4}

\usepackage{color}
\usepackage{graphicx}

\usepackage{dcolumn}
\usepackage{bm}
\usepackage{lscape}
\usepackage{ulem}

\usepackage{txfonts}

\begin{document}
\title{Drastic change in inelastic scattering depending on the development of dineutron correlation in $^{10}$Be}

\author{T. Furumoto}%
\email{furumoto-takenori-py@ynu.ac.jp}
\affiliation{College of Education, Yokohama National University, Yokohama 240-8501, Japan}

\author{T. Suhara}%
\affiliation{Matsue College of Technology, Matsue 690-8518, Japan}

\author{N. Itagaki}%
\affiliation{Yukawa Institute for Theoretical Physics, Kyoto University, Kyoto 606-8502, Japan}

\date{\today}

\begin{abstract}
We investigated the development and breaking of the dineutron correlation in $^{10}$Be by analyzing the elastic and inelastic scatterings with a framework combing the microscopic structure and reaction models.
For studying the structure, the $^{10}$Be nucleus was constructed 
under the assumption of a four-body ($\alpha + \alpha + n + n$) cluster model.
In this work, we focused on the change in the inner structure for the 0$_1^+$, 2$_1^+$, and 2$_2^+$ states when the strength of the spin-orbit interaction is varied.
The inner structure, including various physical quantities such as energy, radius, and transition strength, is drastically influenced by the strength of the spin-orbit interaction.
In particular, the development and breaking of the dineutron correlation is governed by the spin-orbit strength.
The differences in the inner structure can be manifested by applying the obtained wave functions to elastic and inelastic scatterings with a proton target at $E/A =$ 59.4 and 200 MeV.
Although the 0$_1^+$ and 2$_1^+$ states are significantly 
influenced by the spin-orbit strength of the nuclear structure calculation, the elastic and inelastic cross sections are not much affected.
On the other hand, the inelastic cross section of the 2$_2^+$ state depends greatly on the spin-orbit strength of the structure calculation.
Thus, we discovered a way to measure the degree of the development of dineutron cluster structure based on its sensitivity to the inelastic cross section of the 2$_2^+$ state of $^{10}$Be.
\end{abstract}

\maketitle

\section{Introduction}
The cluster structure is one of the key issues in nuclear physics.
In particular, the $\alpha$ cluster plays an important role in imparting an exotic structure in the ground and excited states.
For light systems, especially for the characteristic excited states, $^{8}$Be, $^{12}$C, and $^{16}$O nuclei can be well described with some $\alpha$ clusters.
The most popular state comprising of $\alpha$ clusters is the Hoyle state in $^{12}$C, which is considered to be a dilute triple-$\alpha$ cluster structure~\cite{hoyle1954, uegaki1977, kamimura1981, tohsaki2001, funaki2003}.
Most recently, the algebraic cluster models based on discrete symmetry were successfully applied to describe such nuclear properties~\cite{bijker2014, marin2014, della2017, vitturi2020}.
By adding valence neutrons to such nuclei, a neutron-excess nucleus with a cluster structure can be considered.
Many exotic structures based on $\alpha$ and neutron clusters appear in the ground and excited states.
For example, the halo structure is observed in $^{11}$Li~\cite{tanihata1985}.
It is suggested that the glue-like role of the valence neutrons stabilizes the molecular-orbital structure in Be and C isotopes~\cite{seya1981,itagaki2000-1,itagaki2000-2,itagaki2001,itagaki2002,itagaki2004,suhara2010,suhara2011,fritsch2016,yamaguchi2017}.
In neutron-rich nuclei, since the binding energies of the last neutron are small, neutron-neutron correlation is important.
Thus, the dineutron correlation is considered as the clustering factor in unstable nuclei.
The dineutron model was proposed in Ref.~\cite{migdal1973}, dineutron and diproton structures in various nuclei have been widely investigated and discussed~\cite{zhukov1993,itagaki2002,zhukov2004,hagino2005,hagino2008,enyo2009,aoyama2009,kikuchi2010,spyrou2012,kikuchi2016,lovell2017,oishi2017,hiyama2019,cook2020,kubota2020,wang2021}.

The $^{10}$Be nucleus is a neutron-excess nucleus.
The $^{7}$Be and $^{10}$Be nuclei are produced by the nuclear reaction between cosmic rays and light nuclei (nitrogen and oxygen) in the atmosphere.
These nuclei, especially, $^{10}$Be with a long lifetime, are examined to investigate the activity of the Sun.
The $^{10}$Be nucleus is often considered the typical neutron-rich nucleus with the cluster structure, and in this nucleus, two valence neutrons perform molecular-orbital motions around two $\alpha$ clusters~\cite{seya1981,itagaki2000-1,itagaki2000-2,itagaki2002}.
With regard to its 0$^+$ states, the 0$^+_1$ and 0$^+_3$ states are characterized by the $\pi$ orbit of the valence neutrons around $\alpha$--$\alpha$.
On the other hand, the 0$^+_2$ state has a large $\alpha$--$\alpha$ distance, which is characterized by the $\sigma$ orbit.
Above these 0$^+$ states, rotational bands are formed.
For example, the 2$^+_1$ state is a member of the $K=0$ band together with the ground $0^+$ state and the 2$^+_2$ state is the band head state of the $K=2$ side-band.
This 2$^+_2$ state is located close to the neutron-threshold energy and thus the single-particle motion of the two neutrons and dineutron correlation compete with each other.
Ref.~\cite{itagaki2002} discussed the dineutron correlation and its breaking because of the spin-orbit interaction.
The persistence of the dineutron cluster is sensitive to the strength of the spin-orbit interaction, even if the binding energy of the neutrons from the threshold is kept constant.
The weak spin-orbit interaction favors the dineutron structure.
However, the spin-orbit interaction with realistic strength significantly breaks the dineutron structure.
Exotic features such as the persistence of a dineutron cluster should be identified with experimental data obtained through appropriate nuclear reactions.
In this study, we determined the degree of development of the dineutron correlation in $^{10}$Be from inelastic scattering.

To describe the nuclear reaction part of such analysis, the distorted wave Born approximation (DWBA) and the coupled-channel (CC) calculation are often performed.
Recently, the microscopic construction of the potentials used in the DWBA and CC calculations were developed well to investigate nuclear structures, reactions, and interactions.
Many microscopic descriptions of the nucleon and heavy-ion scatterings are based on the folding model~\cite{sinha1975, brieva1977-1, brieva1977-2, brieva1978, satchler1979, kobos1982, khoa1994, rikus1984, khoa1997, amos2000, khoa2001, furumoto2006, khoa2007, furumoto2008, furumoto2009, furumoto2010, furumoto2016}.
By folding the effective nucleon-nucleon ($NN$) interaction with the projectile and/or target densities, the potential between the projectile and target particles can be obtained.
Reliable and precise information was extracted from experimental data by applying the folding procedure and effective $NN$ interaction~\cite{khoa1997, furumoto2013}.
Recently, the characteristic behaviors of the inelastic cross section attributed to the exotic structure and the property of the potential were predicted~\cite{furumoto2013,sato2019,enyo2019,enyo2020}.

In this study, we investigated the development and breaking of the dineutron correlation in $^{10}$Be for elastic and inelastic scatterings of protons with the framework of the microscopic structure and reaction models.
The present $^{10}$Be nucleus was constructed under the assumption of the four-body ($\alpha + \alpha + n + n$) cluster model.
The stochastic multiconfiguration mixing method was applied to describe as many exotic cluster structures as possible~\cite{ichikawa2011,muta2011}.
The inner structure of $^{10}$Be was artificially modified by changes in the strength of the spin-orbit interaction.
This change in the inner structure resulted in the development of the dineutron correlation and its breaking in $^{10}$Be.
We focused on the 0$_1^+$, 2$_1^+$, and 2$_2^+$ states.
The energies, nuclear size, and expectation values of $< \bm{L} \cdot \bm{S}>$ and $<\bm{S}^2>$ were calculated to investigate the change in the inner structure of the $^{10}$Be nucleus.
To apply the microscopic nuclear reaction model, we obtained the transition density from the wave function.
Specifically, the microscopic coupled channel (MCC) calculation was performed to describe $^{10}$Be elastic and inelastic scatterings by a proton target at $E/A =$ 59.4 and 200 MeV.
The present MCC calculation well reproduces the experimental data for 0$_1^+$ and 2$_1^+$ states at $E/A =$ 59.4 MeV.
Although no experimental data are available for the 2$_2^+$ state, the inelastic cross section is demonstrated in the present MCC calculation.
We discussed the possibility of observing the degree of the development of the dineutron correlation in $^{10}$Be and its breaking for the 2$_2^+$ inelastic cross section.

The reminder of this paper is organized as follows.
In Sec.~\ref{sec:formalism}, we briefly introduce the present structure and reaction models.
In Sec.~\ref{sec:results}, we present the results of the microscopic cluster and reaction models.
The results clarify the drastic changes in the energies, nuclear size, and expectation values.
In addition, the transition strength, transition density, and calculated cross section are discussed.
Lastly, we summarize this paper in Sec.~\ref{sec:summary}.

\section{Formalism}
\label{sec:formalism}

We first constructed several types of the $^{10}$Be nucleus within the 4-body ($\alpha + \alpha + n + n$) cluster model depending on the strength of the spin-orbit interaction.
Subsequently, we constructed the transition density from the wave function.
By applying the transition density to the MCC calculation, we obtained the proton elastic and inelastic cross sections.
The details of the structure and reaction are provided in Refs.~\cite{furumoto2013,furumoto2018,furumoto2021}.
In this section, we introduce the formalism briefly.

\subsection{Microscopic cluster model}
The $^{10}$Be nucleus was constructed by the stochastic multiconfiguration mixing method based on the microscopic cluster model~\cite{ichikawa2011,muta2011}.
The total wave function $\Phi^{J^{\pi} M}$ is expressed by the superposition of the basis states $\Psi_{i}^{J^{\pi} MK}$ as follows:
\begin{equation}
\Phi^{J^{\pi} M} = \sum_{K} \sum_{i} c_{i, K} \Psi_{i}^{J^{\pi} MK}.
\label{mcmwf}
\end{equation}
The eigenstates of the Hamiltonian were obtained by diagonalizing the Hamiltonian matrix.
Additionally, the coefficients, $c_{i, K}$, for the linear combination of Slater determinants were obtained.
We adopted the 500 basis states to obtain the total wave function.
We confirmed that sufficient convergence was achieved with 500 basis states.
In fact, for the 4-body cluster system, even 400 basis states are known to provide sufficient convergence as shown in Ref.~\cite{furumoto2013}.

Next, we introduce various $\alpha + \alpha + n + n$ configurations for the basis states to describe the $^{10}$Be nucleus as follows:
\begin{widetext}
\begin{equation}
\Psi_{i}^{J^{\pi} MK} =   P^{\pi} P^{JMK} {\cal A}
 \big[ \phi_{\alpha} (\bm{r}_{1} \bm{r}_{2} \bm{r}_{3} \bm{r}_{4}, \bm{R}_{1}) \phi_{\alpha} (\bm{r}_{5} \bm{r}_{6} \bm{r}_{7} \bm{r}_{8}, \bm{R}_{2}) 
\phi_{n} (\bm{r}_{9}, \bm{R}_{3}) \phi_{n} (\bm{r}_{10}, \bm{R}_{4}) \big]_{i},
\end{equation}
\end{widetext}
where $\cal A$ is the antisymmetrizer and $\phi_{\alpha}$ and $\phi_{n}$ are the wave functions of $\alpha$ and neutron, respectively.
The wave functions of the $j$-th nucleon, whose spatial coordinate is $\bm{r}_{j}$, is described as a locally shifted Gaussian centered at $\bm{R}$, $\exp[-\nu(\bm{r}_j - \bm{R})^2]$.
Here, the positions of the Gaussian-centered parameter $\bm{R}$ are randomly generated.
The $\alpha$ cluster comprised four nucleons: spin-up proton, spin-down proton, spin-up neutron, and spin-down neutron.
These nucleons shared a common Gaussian-centered parameter, $\bm{R}_{1}$ or $\bm{R}_{2}$.
However, for simplicity, the spin and isospin of each nucleon were not explicitly described in this formula.
The projection onto an eigenstate of parity and angular momentum by
operators $P^{\pi}$ and $P^{JMK}$ was performed numerically. 
For the Euler angle integral, the number of mesh points was $16 \times 24 \times 16$, i.e., $6144$.
The value of $M$ represents the $z$ component of the angular momentum in the laboratory frame.
The energy does not depend on $M$; however, it depends on $K$, which is the $z$ component of the angular momentum in the body-fixed frame.

The Hamiltonian is same as in Refs.~\cite{furumoto2013,furumoto2018}.
The two-body interaction includes the central, spin-orbit, and Coulomb parts.
The Volkov No.2 effective potential was applied to the central part~\cite{volkov1965}.
We used the same parameter set including $\nu$ ($\nu$ is the width parameter of the wave function) as in Refs.~\cite{furumoto2013,furumoto2018}.

Here, we introduce the spin-orbit term of the G3RS potential~\cite{tamagaki1968,yamaguchi1979},
\begin{equation}
V= V_{\rm LS}( e^{-d_{1}r^{2}}-e^{-d_{2}r^{2}}) P(^{3}O) \bm{L} \cdot {\bm{S}},
\end{equation}
where operator $\bm{L}$ represents the relative angular momentum, and $\bm{S}$ represents the spin ($\bm{S}_1+\bm{S}_2$).
$P(^{3}O)$ is the projection operator onto the triplet odd state.
The parameters of $d_1$ and $d_2$ are same as in Refs.~\cite{furumoto2013,furumoto2018}.
In this study, the strength of the spin-orbit interaction, $V_{\rm LS}$, is changed by hand.
This value is often fixed at about 2000 MeV to reproduce the data of the $^{10}$Be nucleus.
In this study we set this value as 0, 500, 1000, 1500, 2000, 2500, 3000, 3500, and 4000 MeV.
The effect of changing the $V_{\rm LS}$ value will be discussed in the next sections.

To connect the nuclear structure and reaction calculations, we prepared the diagonal and transition densities in the same manner as in Ref.~\cite{baye1994}.

\subsection{MCC model}
After calculating the diagonal and transition densities, we performed the nuclear reaction calculation.
We applied the calculated transition densities to MCC calculations with the complex $G$-matrix interaction MPa~\cite{yamamoto2014,yamamoto2016}.
The MPa interaction has been successful applied for nuclear reactions~\cite{furumoto2016,qu2017,furumoto2019,furumoto2021}.
The detailed calculation procedure for the folding potential is described in literature~\cite{khoa2002,furumoto2008,furumoto2021}; hence, only the essence of the MCC calculation is briefly introduced in this paper.

The diagonal and coupling potentials are necessary to solve the CC calculation.
The potentials including the spin-orbit part are constructed by a folding procedure based on Refs.~\cite{khoa2002,furumoto2008,furumoto2021}.
The central direct $U^{\rm (CE)}_{D}$ and exchange $U^{\rm (CE)}_{\rm EX}$ potentials are simply described as follows:
\begin{widetext}
\begin{eqnarray}
U^{\rm (CE)}_{D}(R; E/A) &=& \int{ \rho_{tr}(r) v^{\rm (CE)}_{D}(\bm{s}, k_{F}; E/A) d\bm{r} }, \label{eq:cent-direct}\\
U^{\rm (CE)}_{\rm EX}(R; E/A) &=& \int{ \left( \frac{R}{x} \right)^{\lambda} \rho_{tr}(x) \frac{3}{k^{\rm (eff)}_F s} j_1(k^{\rm (eff)}_F s) v^{\rm (CE)}_{\rm EX}(\bm{s}, k_{F}; E/A) j_0 (ks) d\bm{r} }, \label{eq:cent-exchange}
\end{eqnarray}
\end{widetext}
where $R$ is the radial distance between the incident $^{10}$Be nucleus and the target proton.
$E/A$ is the incident energy per nucleon.
$\lambda$ indicates the multipolarity; $\rho_{tr}$ is the transition density; $s$ is the radial distance between a nucleon in the projectile nucleus and the target proton.
Further, $\bm{s} = \bm{r} - \bm{R}$.
$\bm{x} = \frac{1}{2}(\bm{r} + \bm{R})$.
$k_F$ is the Fermi momentum derived from the densities of the initial and final states, and $j_0$ and $j_1$ are the spherical Bessel function of rank 0 and 1, respectively.
$k_F^{\rm (eff)}$ is the effective Fermi momentum defined in Ref.~\cite{campi1978}, and $v^{\rm (CE)}_D$ and $v^{\rm (CE)}_{\rm EX}$ are the complex $G$-matrix interactions for the central direct and exchange terms, respectively.
We note that the descriptions of Eqs.~(\ref{eq:cent-direct}) and (\ref{eq:cent-exchange}) are simplified.
The proton and neutron densities are separately folded.
The Coulomb potential is also obtained by folding with the nucleon-nucleon Coulomb interaction and proton density.
After folding calculation, we obtained the central part of the potential by combining the direct and exchange potentials as follows:
\begin{equation}
U^{\rm (CE)} = U^{\rm (CE)}_D + U^{\rm (CE)}_{\rm EX}.
\end{equation}

The diagonal and coupling potentials for the spin-orbit part were obtained in the same manner as described in Refs.~\cite{furumoto2008,furumoto2021}, as follows:
\begin{widetext}
\begin{eqnarray}
U^{\rm (LS)}_D(R; E/A)&=&\frac{1}{4R^2} \int{\bm{R}\cdot (\bm{R}-\bm{r}) \rho_{tr}(\bm{r}) v^{\rm (LS)}_D(\bm{s}, k_F; E/A) d\bm{r}}, \label{eq:LS-direct}\\
U^{\rm (LS)}_{\rm EX}(R; E/A)&=& \pi \int{ds s^3\left[ \frac{2j_0(ks)}{R}\rho_1 (R, s; E/A) + \frac{j_1(ks)}{2k}\delta_0 (R, s; E/A) \right] }, \label{eq:LS-exchange}
\end{eqnarray}
where,
\begin{eqnarray}
\delta_0(R, s; E/A)&=&\frac{1}{2}\int^{+1}_{-1} {dq \frac{v^{\rm (LS)}_{\rm EX}(s, k_{F}; E/A)}{X} \bigg{[} \frac{3}{k^{\rm{eff}}_{F} s} j_1(k^{\rm{eff}}_{F} s) \frac{d}{dx} \rho_{tr}(x)\bigg|_{x=X}} + s\rho_{tr} (X) \frac{d}{dx}k^{\rm{eff}}_{F}(x)\bigg|_{x=X} \frac{d}{dy}\left( \frac{3}{y}j_1(y)\right) \bigg|_{y=k^{\rm{eff}}_{F} s} \bigg] , \label{eq:LS-exchange0}\\
\rho_1(R, s; E/A) &=&\frac{1}{2}\int^{+1}_{-1}dq q v^{\rm (LS)}_{\rm EX}(s, k_{F}; E/A) \frac{3}{k^{\rm{eff}}_{F} s} j_1(k^{\rm{eff}}_{F} s)\rho_{tr} (X), \label{eq:LS-exchange1}
\end{eqnarray}
\end{widetext}
where $X=\sqrt{R^2+s^2/4+Rsq}$.
$v^{\rm (LS)}_D$ and $v^{\rm (LS)}_{\rm EX}$ are the complex $G$-matrix interactions for the direct and exchange terms for the spin-orbit interaction, respectively.
The descriptions of Eqs.~(\ref{eq:LS-direct}), (\ref{eq:LS-exchange0}), and (\ref{eq:LS-exchange1}) are simplified.
In fact, the proton and neutron densities are separately folded.
We also obtained the spin-orbit part of the potential by combining the direct and exchange potentials as
\begin{equation}
U^{\rm (LS)} \bm{\ell} \cdot \bm{\sigma} = (U^{\rm (LS)}_D + U^{\rm (LS)}_{\rm EX}) \bm{\ell} \cdot \bm{\sigma}.
\end{equation}

After obtaining the folding potential, we modified the strength of the imaginary part of the folding potential.
Since the complex $G$-matrix is constructed with infinite nuclear matter, the strength of the imaginary part is often adjusted for the finite nucleus because these level densities are quite different.
Therefore, we consider the incident-energy-dependent renormalization factor, $N_W = 0.5 + (E/A)/1000$~\cite{furumoto2019}, for the imaginary part of the folding model potential.
Concretely, the potentials are modified as
\begin{eqnarray}
U&=&U^{\rm (CE)}+U^{\rm (LS)} \bm{\ell} \cdot \bm{\sigma} \\
&=&V^{\rm (CE)} +i W^{\rm (CE)} + (V^{\rm (LS)} + i W^{\rm (LS)}) \bm{\ell} \cdot \bm{\sigma} \\
&\to&V^{\rm (CE)} +i N_W W^{\rm (CE)} + (V^{\rm (LS)} + i N_W W^{\rm (LS)}) \bm{\ell} \cdot \bm{\sigma},
\end{eqnarray}
where $V^{\rm (CE)}$, $W^{\rm (CE)}$, $V^{\rm (LS)}$, and $W^{\rm (LS)}$ are the central real, central imaginary, spin-orbit real, and spin-orbit imaginary potentials, respectively.
In other words, in this study, we did not use any additional parameter to calculate the $^{10}$Be scatterings by the proton target.

After fixing the central and spin-orbit potentials, the scattering matrix dependent on the total angular momentum was obtained from the folded potentials by solving the CC equation based on the Stormer method.
Relativistic kinematics was used in the calculation.
The cross section was calculated with the scattering amplitude derived from the scattering matrix as shown in Ref.~\cite{DNR}.

\section{Results}
\label{sec:results}
The calculation results obtained using the microscopic nuclear structure and reaction models are described in this section.
First, the results of the microscopic cluster model are introduced.
By changing the strength of the spin-orbit interaction, various changes were indued in the inner structures and physical quantities such as energies, radii, and expectation values of $< \bm{L} \cdot \bm{S}>$ and $< \bm{S}^2 >$ of the 0$^+_1$, 2$^+_1$, and 2$^+_2$ states.
Using these results, we discuss the degree of the development of the dineutron correlation in $^{10}$Be.
Next, the effect of the development of the dineutron correlation on the elastic and inelastic cross sections is described.

\subsection{Change in structure information}

\begin{figure}[h]
\centering
\includegraphics[width=8cm]{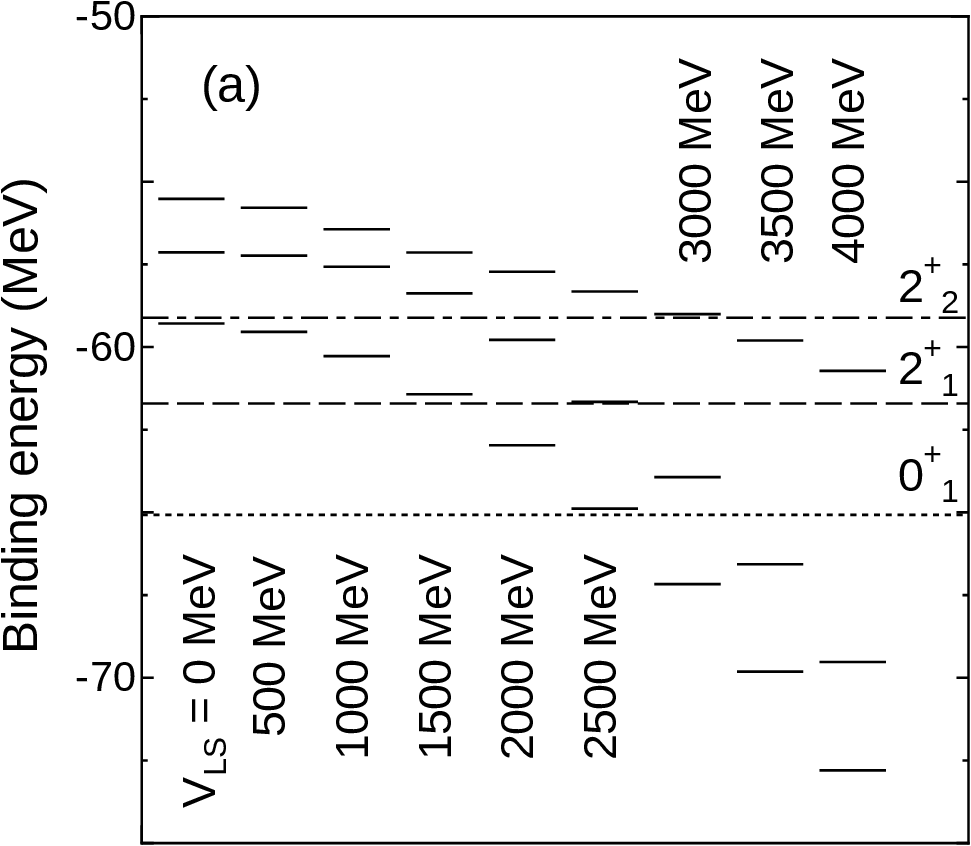} \\
\includegraphics[width=8cm]{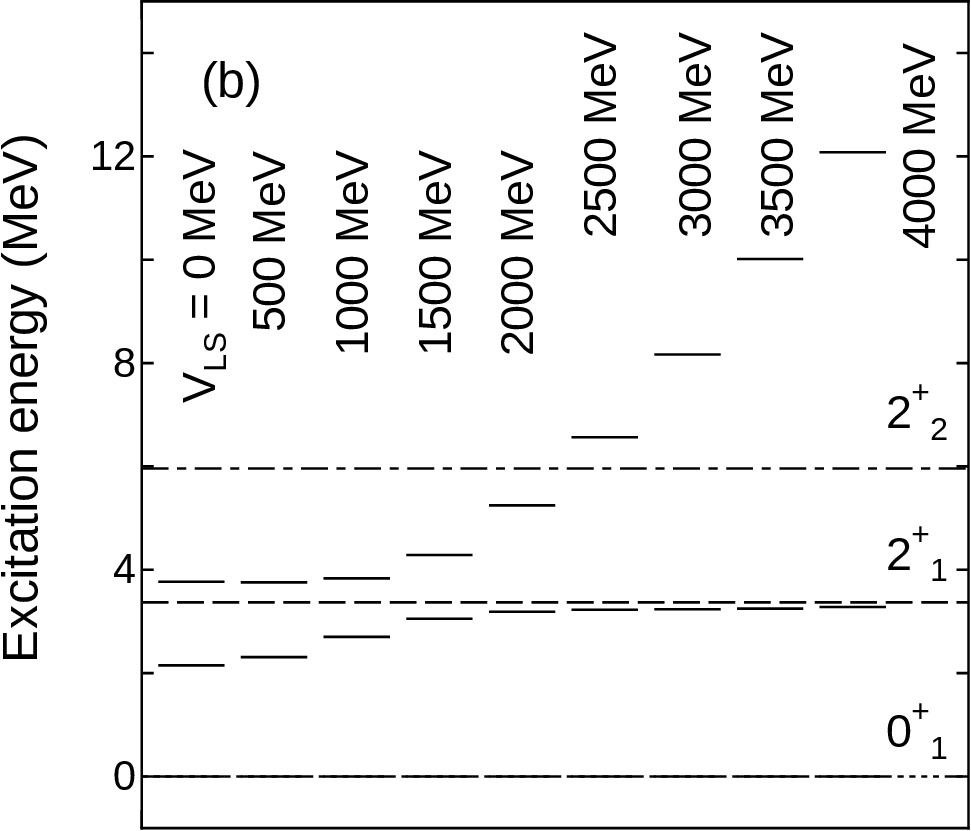}
\caption{\label{fig:energy}
(a) Binding energy and (b) excitation energy for the 0$^+_1$, 2$^+_1$, and 2$^+_2$ states of $^{10}$Be with $V_{\rm LS}$ = 0--4000 MeV.
The dotted, dashed, and dot-dashed lines are the experimental data from \cite{NNDC}.}
\end{figure}

Figure~\ref{fig:energy} shows the calculated (a) binding and (b) excitation energies compared with the experimental data.
By changing the strength $V_{\rm LS}$ in the range 0--4000 MeV, the binding energy of the 0$_1^+$ state was drastically changed.
A large $V_{\rm LS}$ value results in strong binding energy.
On the other hand, a small $V_{\rm LS}$ value results in weak binding energy.
The 2$^+_1$ state with the strong $V_{\rm LS}$ value also gives the strong binding energy.
The $2^+_2$ state gains binding energy gently compared to the other two states because one of the valence neutrons is exited from the spin-orbit favored orbit to unfavored one.
For better understanding, we show the excitation energy in Fig.~\ref{fig:energy}.
The excitation energy of the 2$^+_1$ state approaches the experimental value with increase in the strength of the spin-orbit interaction.
On the other hand, the excitation energy of the 2$^+_2$ state rapidly jumps up with an increase in the strength of the spin-orbit interaction.
The calculated values were almost identical to the experimental data around 2000 MeV.
The separation between the 2$^+_1$ and 2$^+_2$ states is caused by the spin-orbit splitting of the $p_{3/2}$ and $p_{1/2}$ single-particle orbits for the last neutron.

\begin{figure}[h]
\centering
\includegraphics[width=6.4cm]{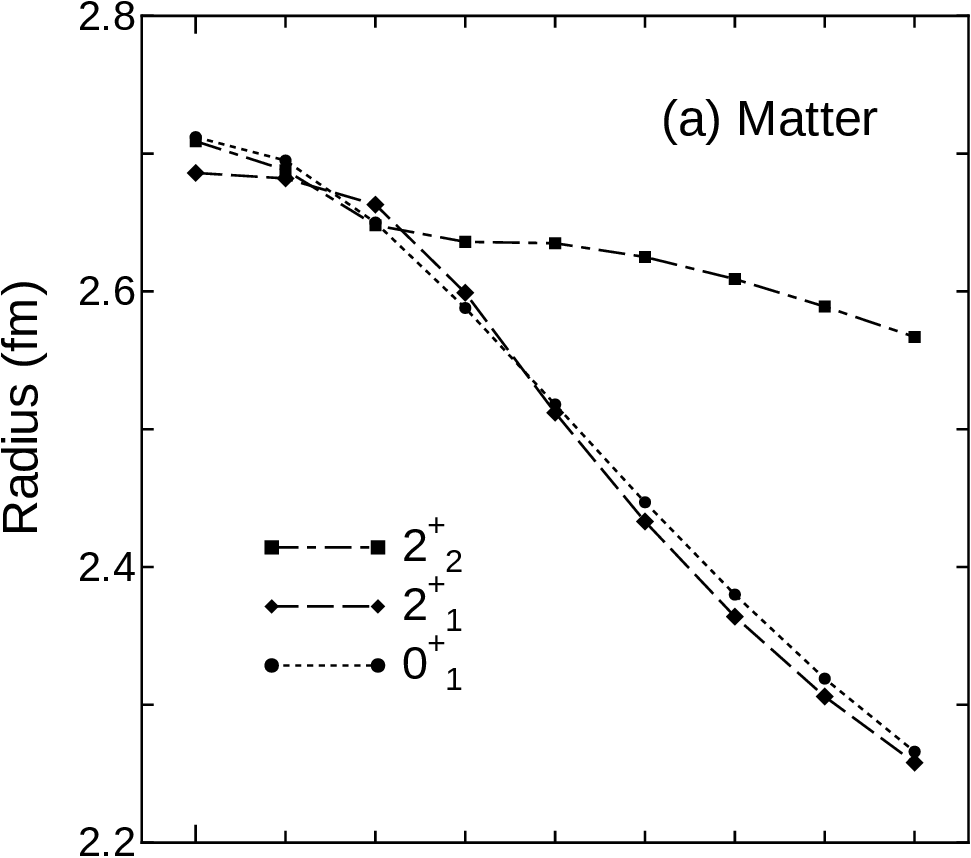} \\
\includegraphics[width=6.4cm]{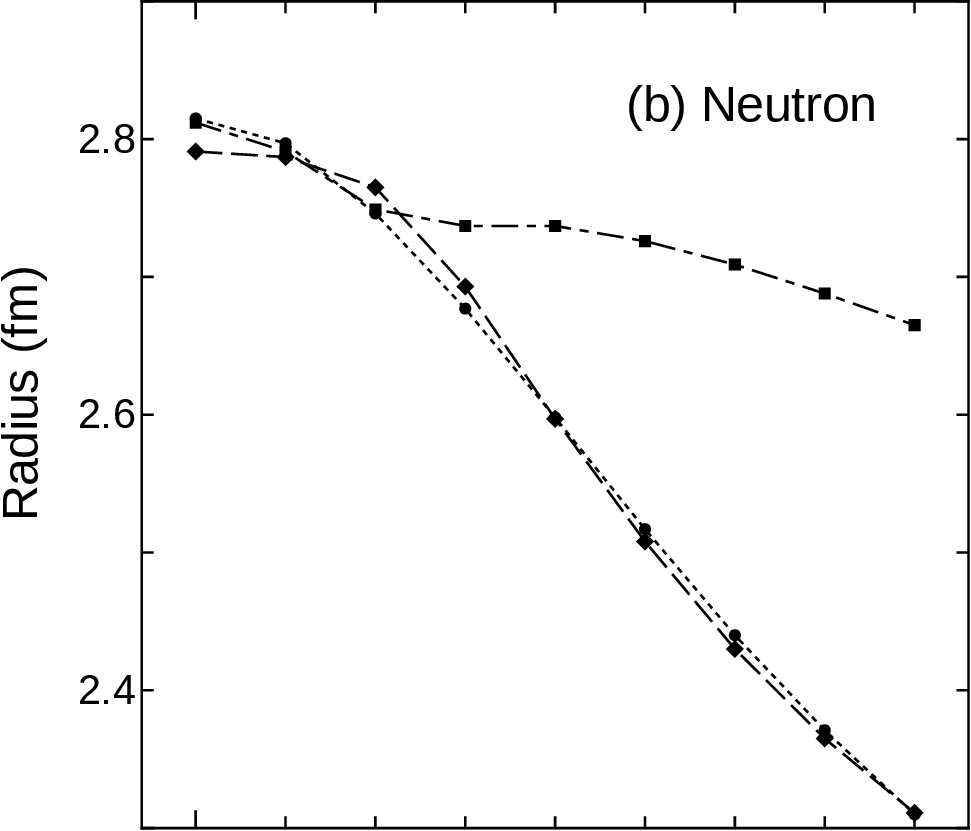} \\
\includegraphics[width=6.4cm]{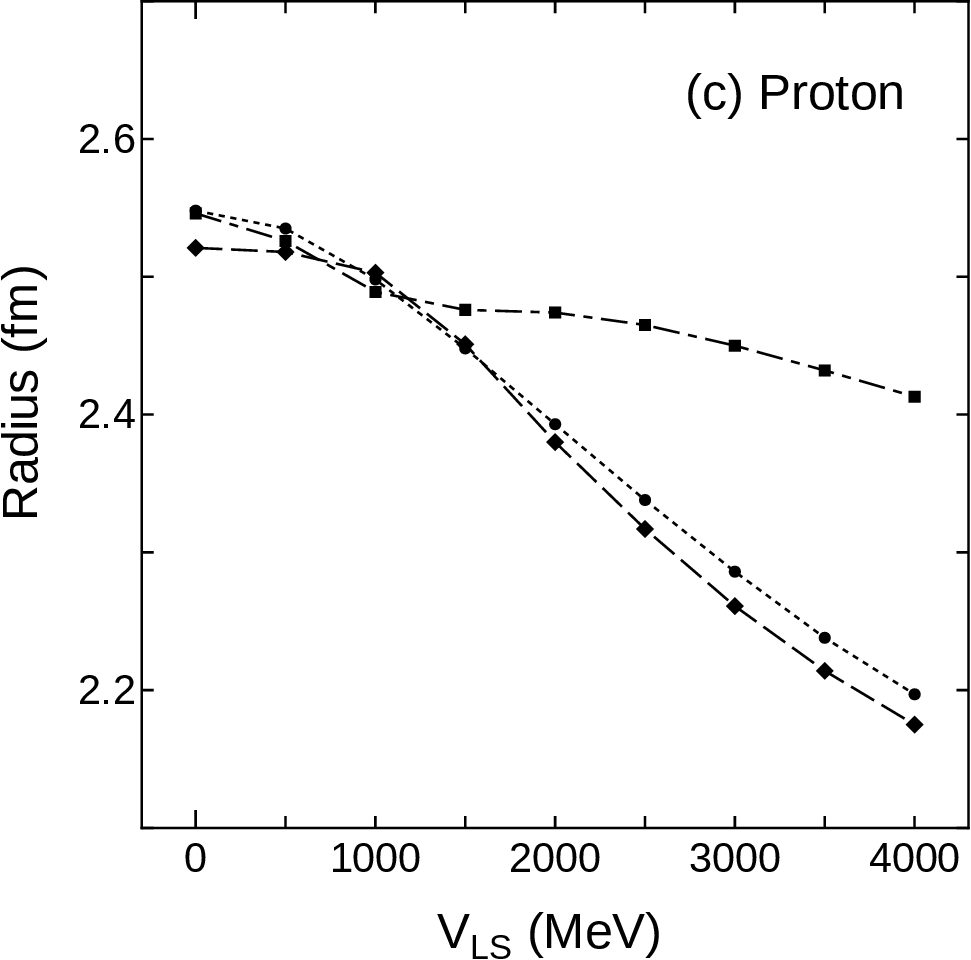}
\caption{\label{fig:radius}
The root mean squared radius for (a) point nucleon matter, (b) point neutron, and (c) point proton with $V_{\rm LS}$ = 0--4000 MeV.
The dotted, dashed, dot-dashed curves are the results for the 0$^+_1$, 2$^+_1$, and 2$^+_2$ states, respectively.}
\end{figure}

Figure~\ref{fig:radius} shows the calculated root mean squared radii for the (a) point nucleon matter, (b) point neutron, and (c) point proton of the 0$^+_1$, 2$^+_1$, and 2$^+_2$ states with $V_{\rm LS}$ = 0--4000 MeV.
The dotted, dashed, dot-dashed curves are the results for the 0$^+_1$, 2$^+_1$, and 2$^+_2$ states, respectively.
The experimental values of the proton and matter radii of the $^{10}$Be nucleus were 2.357 $\pm$ 0.018 fm~\cite{nortershauser2009} and 2.39 $\pm$ 0.02 fm~\cite{tanihata1985}, respectively.
The calculated data was very close to the experimental data around $V_{\rm LS}$ = 2500 MeV.
The increase in the strength of the spin-orbit interaction is responsible for the small size for the point nucleon-matter, neutron, and proton because of the strong binding, as shown in Fig.~\ref{fig:energy}.
The size of the 0$^+_1$ and 2$^+_1$  states show behaviors consistent with the strong binding.
On the other hand, the 2$^+_2$ state, which is larger than the 0$^+_1$ and 2$^+_1$ states except for very small $V_{\rm LS}$ values, exhibits much gentler shrinkage behavior with increasing binding energy.
This result implies that the structure of the 2$^+_2$ state is quite different from that of the ground and 2$^+_1$ states, when the strong spin-orbit interaction is applied.

\begin{figure}[h]
\centering
\includegraphics[width=8cm]{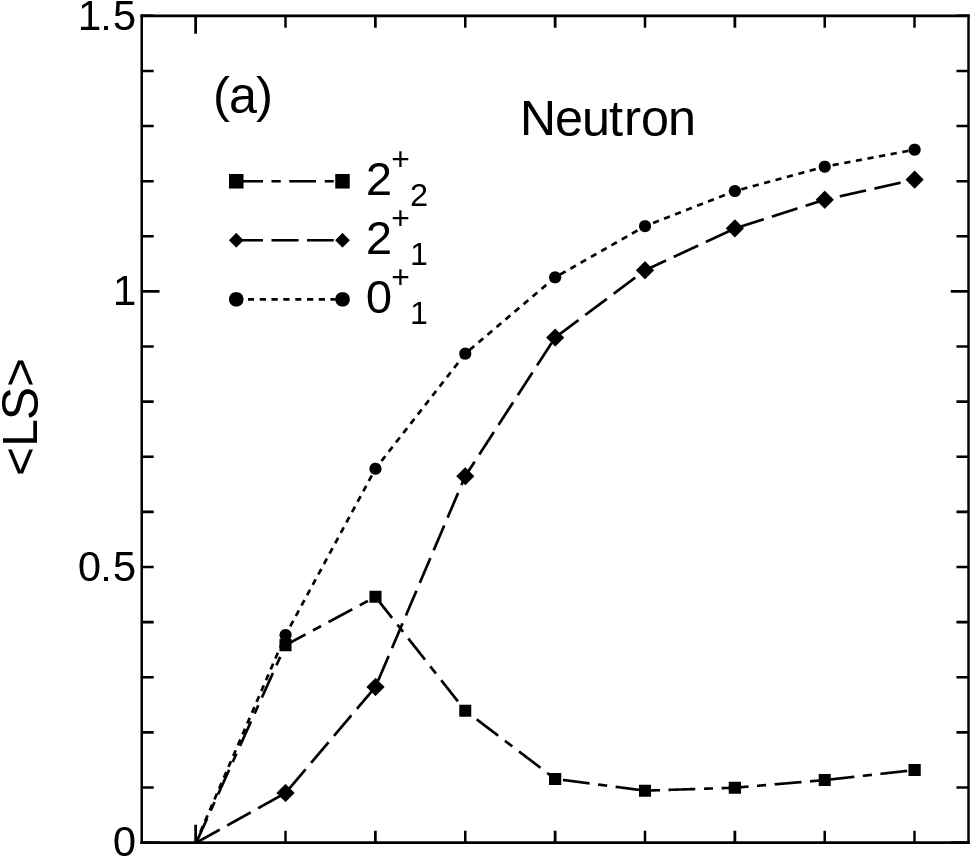} \\
\includegraphics[width=8cm]{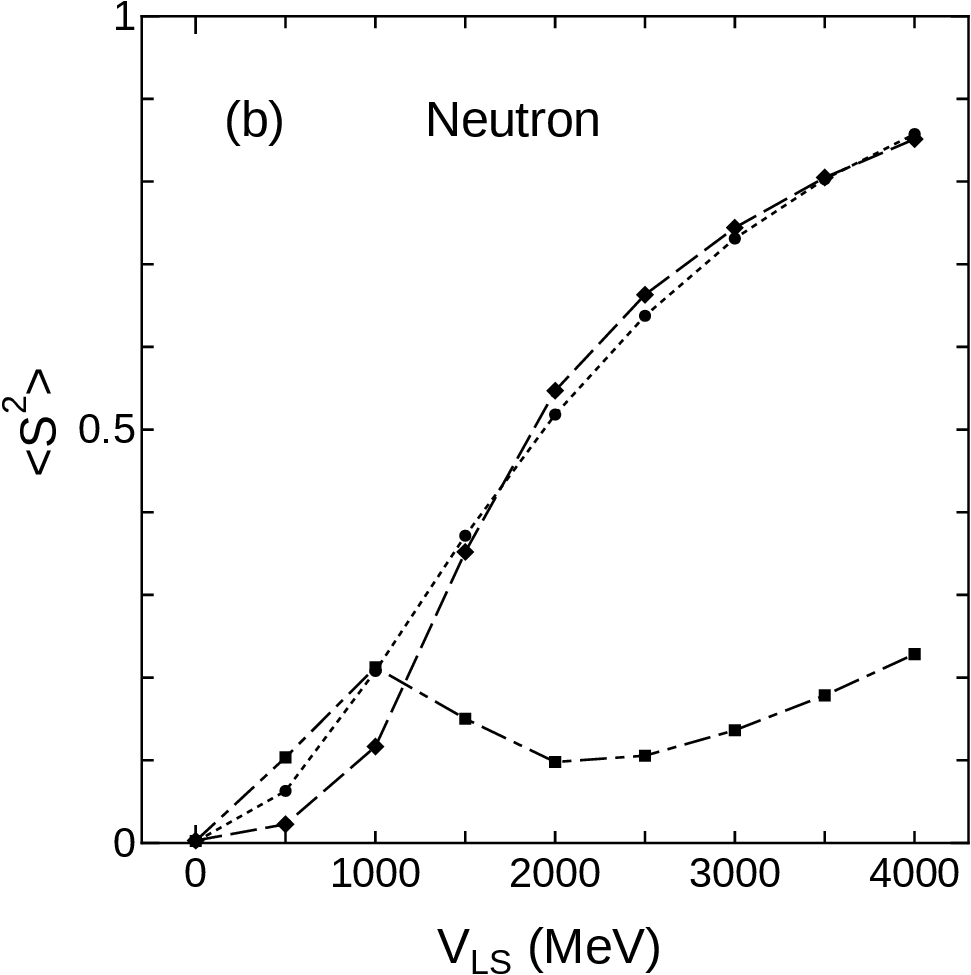}
\caption{\label{fig:LSandSS}
Expectation value of $< \bm{L} \cdot \bm{S}>$ and $< \bm{S}^2 >$ with $V_{\rm LS}$ = 0--4000 MeV.
}
\end{figure}

For more precise analysis, we investigated other structural information.
Figure~\ref{fig:LSandSS} shows the expectation values of (a) $< \bm{L} \cdot \bm{S}>$ and (b) $< \bm{S}^2 >$ for the neutron with $V_{\rm LS}$ = 0--4000 MeV.
The expectation values of the proton part are 0 because all protons are in the $\alpha$ cluster.
The 0$^+_1$ and 2$^+_1$ states clearly exhibit similar behaviors during the growth of the expectation value with changing strength of the spin-orbit interaction.
Thus, the $0^+_1$ and $2^+_1$ states have similar inner structures and undergo similar change with increasing strength of the spin-orbit interaction.
The expectation value of the 2$^+_2$ state shows different behavior unlike the behavior of the $0^+_1$ and $2^+_1$ states.
This finding implies that the inner structure of the 2$^+_2$ state is different from the structures of the 0$^+_1$ and 2$^+_1$ states when the strong spin-orbit interaction is applied.

Now, we describe the structure in the case of weak spin-orbit interaction.
In the range of $V_{\rm LS}$ = 0--1000 MeV, the expectation values of all states become close to each other.
The size of the present states in $^{10}$Be is also comparable to each other with the weak spin-orbit interaction as shown in Fig.~\ref{fig:radius}.
However, the binding and/or excitation energies are different.
$^{10}$Be has a 2$\alpha$+dineutron structure with weak spin-orbit interaction as mentioned in Ref.~\cite{itagaki2002}.
All clusters, two $\alpha$s and dineutron, are spin saturated, and therefore, $< \bm{L} \cdot \bm{S}>$, and $< \bm{S}^2 >$ are 0, as shown in Fig.~\ref{fig:LSandSS}.
The 0$^+_1$ and 2$^+_1$ states are composed of the $K=0$ component, while the 2$^+_2$ state is composed of the $K=2$ component.
In fact, all the states have the same inner structure, various physical quantities, radii, $< \bm{L} \cdot \bm{S}>$, and $< \bm{S}^2 >$, and hence, they show the same value as reconfirmed in this work.

With the strong spin-orbit interaction, the dineutron is broken and neutrons occupy the molecular orbits, as observed from the large spin-orbit splitting (Fig.~\ref{fig:energy}) and large expectation values of $< \bm{L} \cdot \bm{S}>$, and $< \bm{S}^2 >$ with the strong spin-orbit interaction.
The excess neutrons in the 0$^+_1$ and 2$^+_1$ states occupy the ($\pi_{3/2}$)$^2$ orbitals around 2$\alpha$ clusters, while the excess neutrons in the 2$^+_2$ state occupy the ($\pi_{3/2}$)($\pi_{1/2}$) orbitals.
Because the $\pi_{1/2}$ orbital is a spin-orbit-unfavored one, it tends to leave from the $\alpha$ clusters.
Therefore, the radius of the 2$^+_2$ state is larger than the radii of the 0$^+_1$ and 2$^+_1$ states as shown in Fig.~\ref{fig:radius}.

\begin{figure}[h]
\centering
\includegraphics[width=8cm]{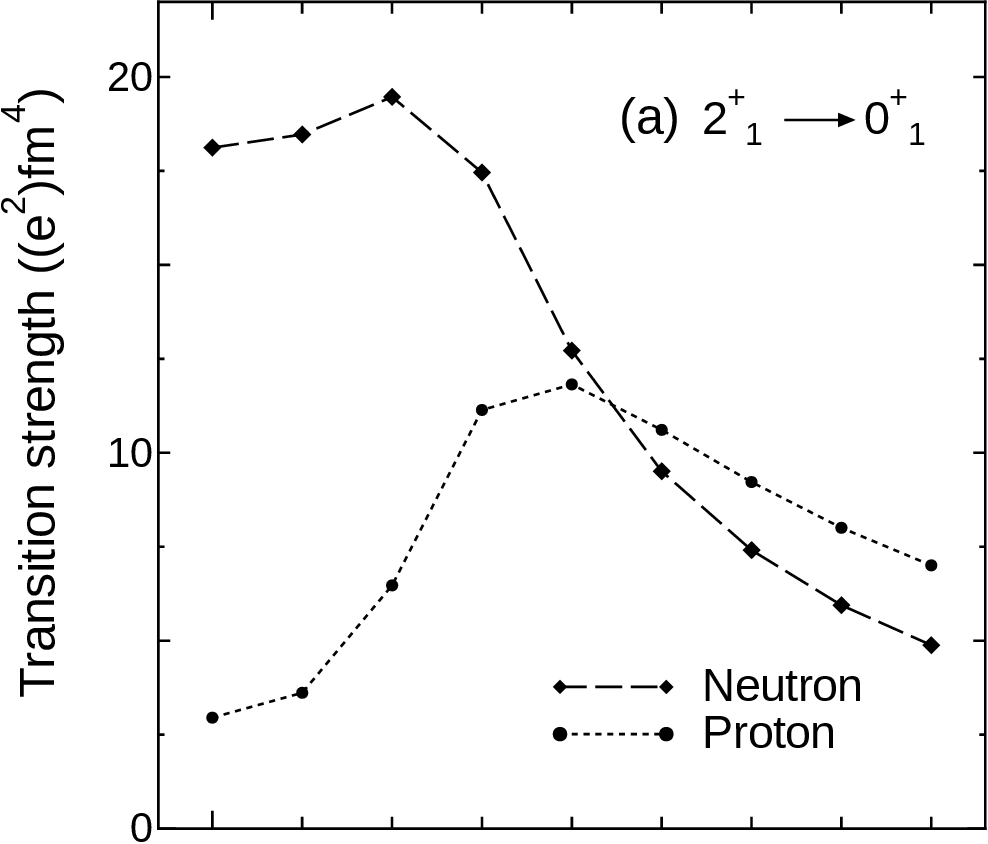} \\
\includegraphics[width=8cm]{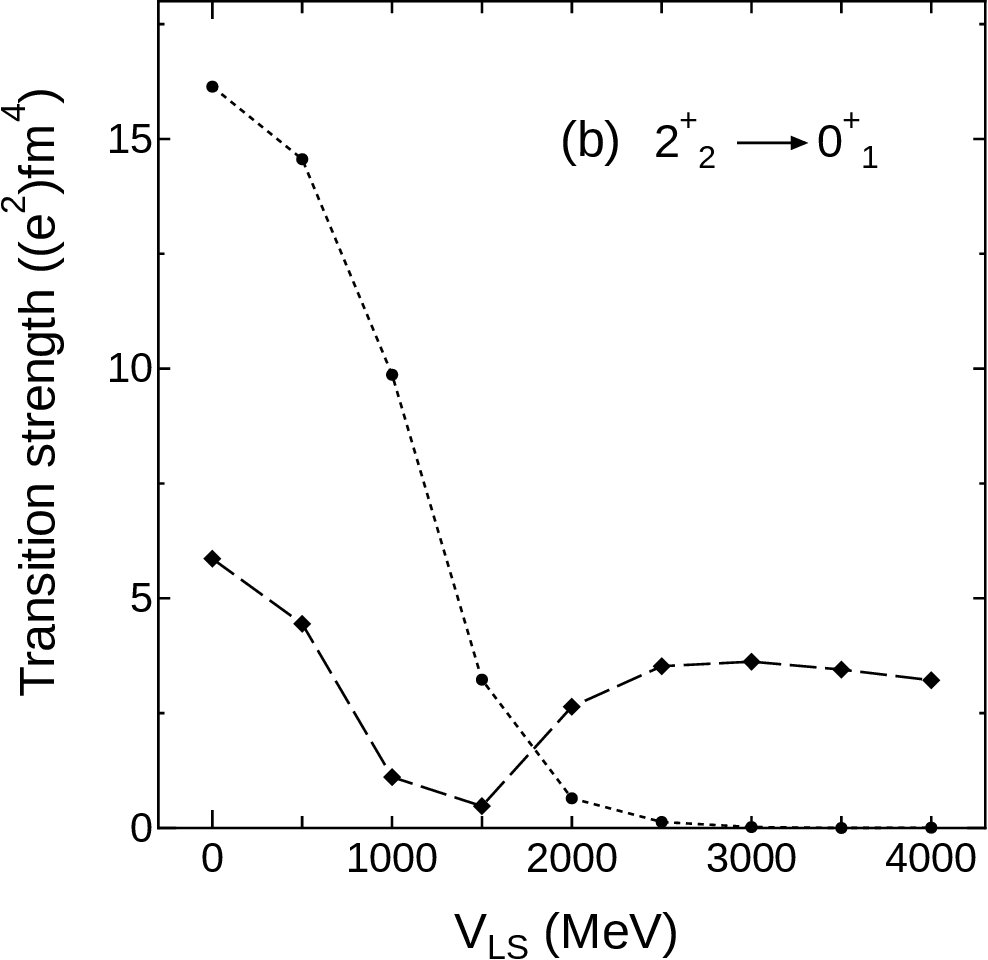}
\caption{\label{fig:trst}
Transition strength ($B(E2)$ and that for the neutron part) (a) from the 2$^+_1$ state to the 0$^+_1$ state and (b) from the 2$^+_{2}$ state to the 0$^+_1$ state with $V_{\rm LS}$ = 0--4000 MeV.
The dotted and dashed lines represent the values for the proton and neutron, respectively.}
\end{figure}

Next, we investigated the transition strength and probability of the ground and excited states.
Figure~\ref{fig:trst} shows the transition strengths ($B(E2)$ and the transition strength for the neutron part) from (a) the 2$^+_1$ state and (b) the 2$^+_2$ state to the 0$^+_1$ state with $V_{\rm LS}$ = 0--4000 MeV.
Although similar values were obtained for the 0$^+_1$ and 2$^+_1$ states as shown in Figs.~\ref{fig:radius} and \ref{fig:LSandSS}, the change in the transition strength from the 2$^+_1$ state to the 0$^+_1$ state is not simple.
Under weak spin-orbit interaction, the strengths of the proton and neutron parts are quite different.
On increasing the $V_{\rm LS}$ value, the transition strength of the proton part becomes large.
For $V_{\rm LS}$ = 2000--4000 MeV, both the proton and neutron parts decrease together.
This behavior can be understood by observing the transition density.
The changes in the behaviours of the transition strengths of the proton and neutron parts from the 2$^+_2$ state to the 0$^+_1$ state with changes in the strength of the spin-orbit interaction is seen in Fig.~\ref{fig:trst}.
The changes in the transition strength from the 2$^+_2$ state to the ground state for the neutron part are caused by the sign reversal of the transition density.
This finding can be confirmed from the transition density.

\begin{figure}[h]
\centering
\includegraphics[width=6cm]{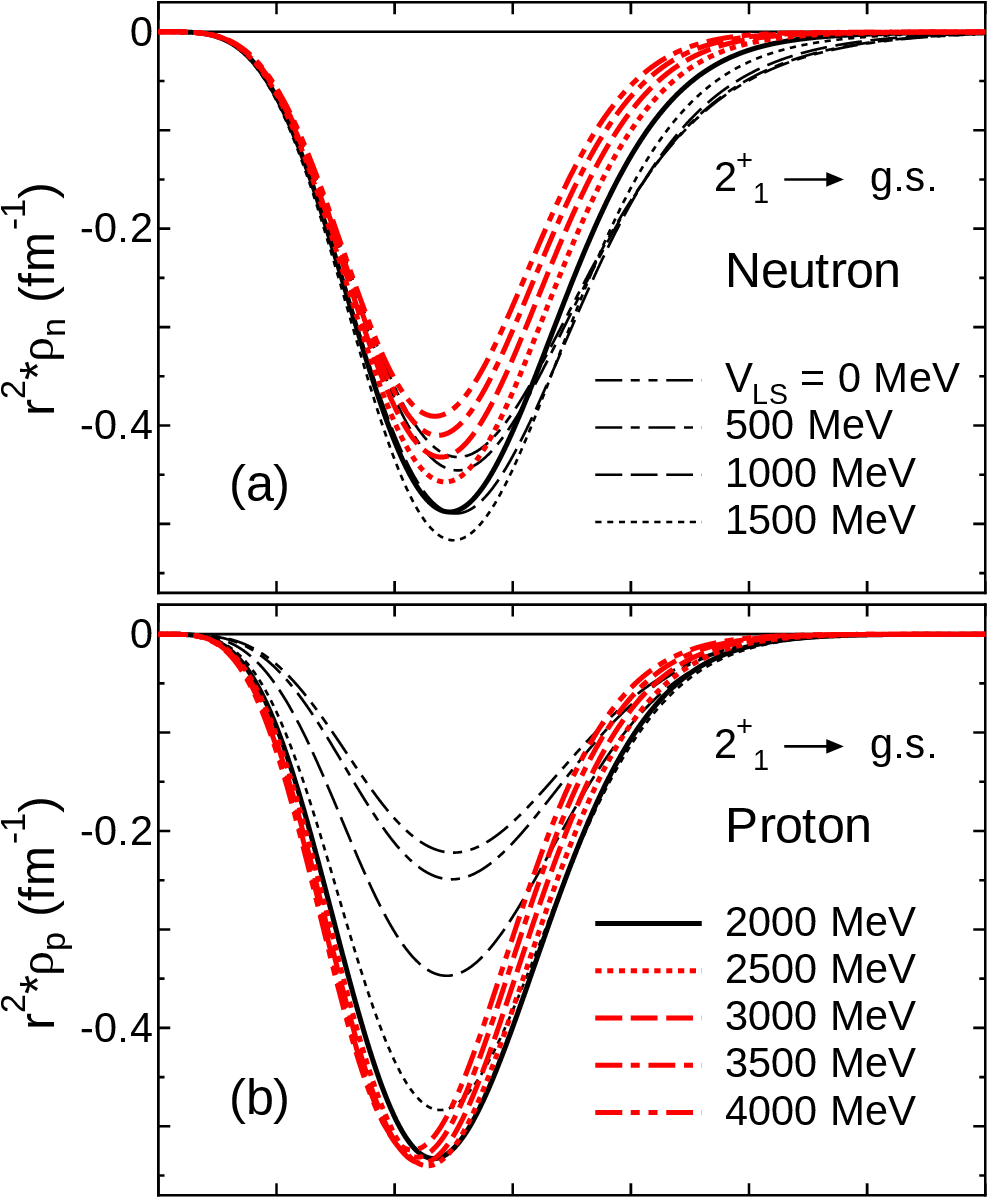} \\
\includegraphics[width=6cm]{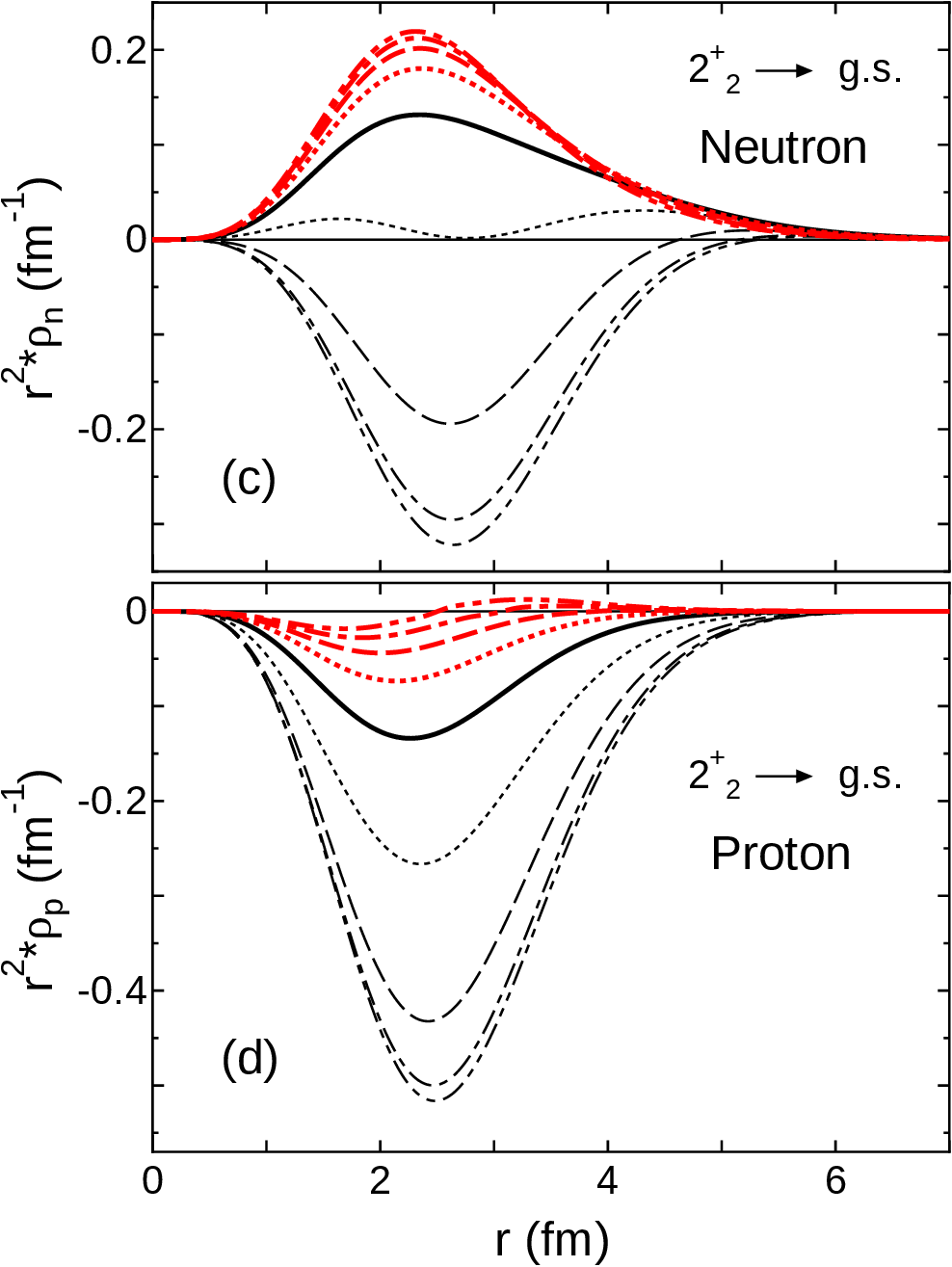}
\caption{\label{fig:trden}
Transition density (a) from the 2$^+_1$ state to the 0$^+_1$ state for the neutron part, (b) from the 2$^+_1$ state to the 0$^+_1$ state for the proton part, (c) from the 2$^+_2$ state to the 0$^+_1$ state for the neutron part, and (d) from the 2$^+_2$ state to the 0$^+_1$ state for the proton part.
The 2-dots-dashed, dot-dashed, dashed, dotted, bold-solid, (red) bold-dotted, (red) bold-dashed, (red) bold-dot-dashed, and (red) bold-2-dots-dashed curves are the results with $V_{\rm LS}$ = 0, 500, 1000, 1500, 2000, 2500, 3000, 3500, and 4000 MeV, respectively.}
\end{figure}

To understand the behavior of the transition strength, we examine the distribution of the transition density in Fig.~\ref{fig:trden}.
Figure~\ref{fig:trden} shows the distribution of the transition density (a) from the 2$^+_1$ state to the 0$^+_1$ state for the neutron part, (b) from the 2$^+_1$ state to the 0$^+_1$ state for the proton part, (c) from the 2$^+_2$ state to the 0$^+_1$ state for the neutron part, and (d) from the 2$^+_2$ state to the 0$^+_1$ state for the proton part.
The transition density from the 2$^+_1$ state to the 0$^+_1$ state for the proton part evolves with the strength as $V_{\rm LS}$ varies in the range of 0 --2000 MeV.
For $V_{\rm LS}$ = 2000--4000 MeV, the distribution of the transition density shifts to the inner part.
This behavior results in a change in the transition strength, as shown in Fig.~\ref{fig:trst}.
However, for transition from the 2$^+_2$ state to the 0$^+_1$ state for the neutron part, the sign of the transition density is reversed with changes in the strength of the spin-orbit interaction.
The sizes of the 0$^+_1$, 2$^+_1$, and 2$^+_2$ states exhibit simple behaviors with changes in the strength of the spin-orbit interaction, as shown in Fig.~\ref{fig:radius}.
However, the transition strength and transition density exhibit complicated behaviors with changes in the strength of the spin-orbit interaction.
These results indicate that changes in the transition density are affected by the changes in not only the nuclear size but also the inner structure.

\subsection{Effect on cross sections}

Next, we used the transition density for MCC calculation.
The effect of changes in the inner structure on the elastic and inelastic cross sections was investigated.
The input value of the excitation energy in the MCC calculation is based on experimental data.

\begin{figure}[h]
\centering
\includegraphics[width=8cm]{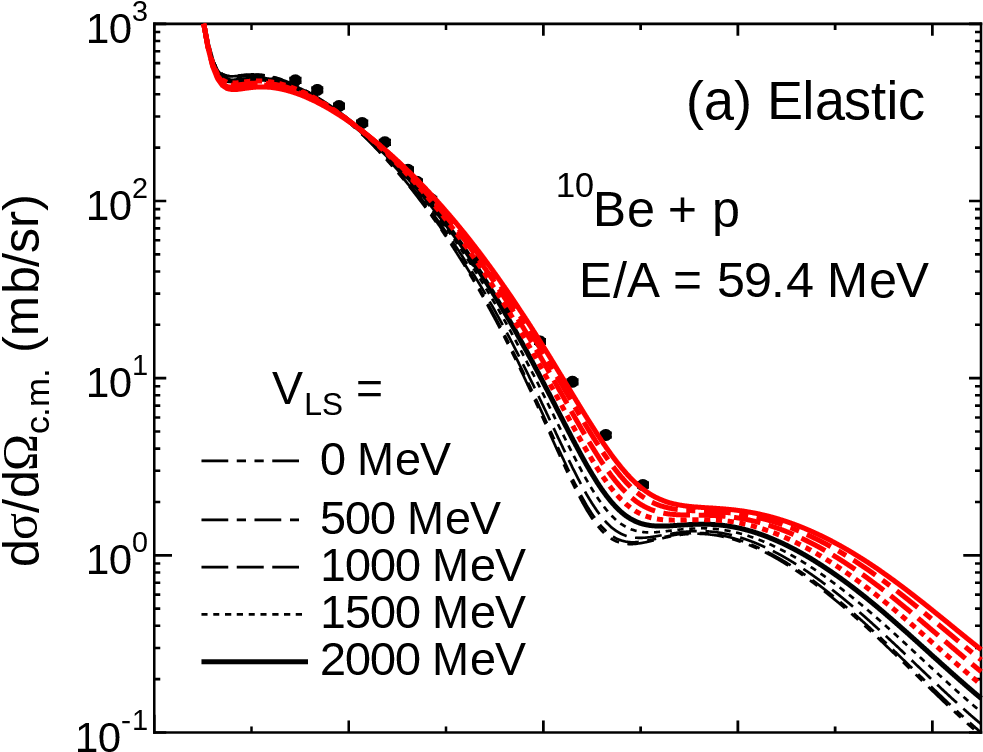}\\
\includegraphics[width=8cm]{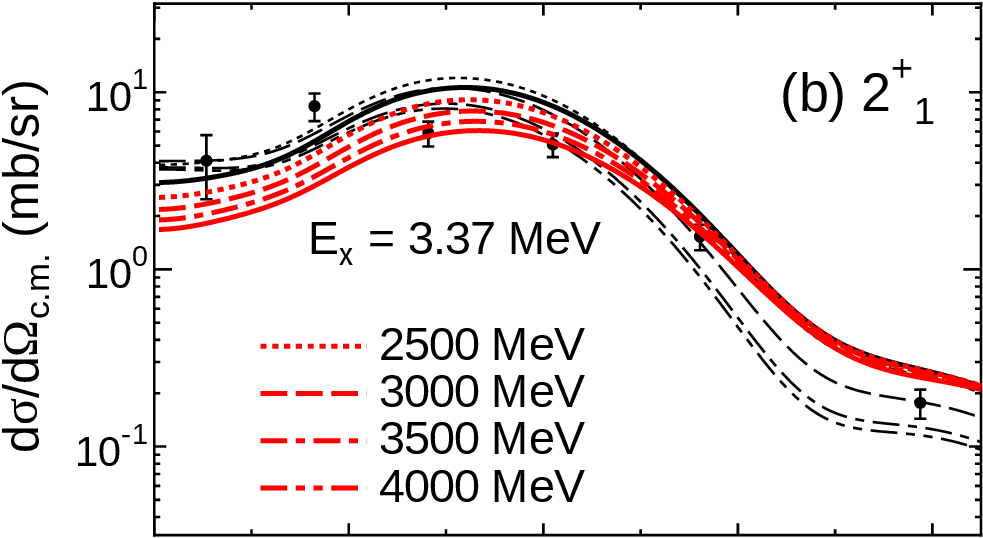}\\
\includegraphics[width=8cm]{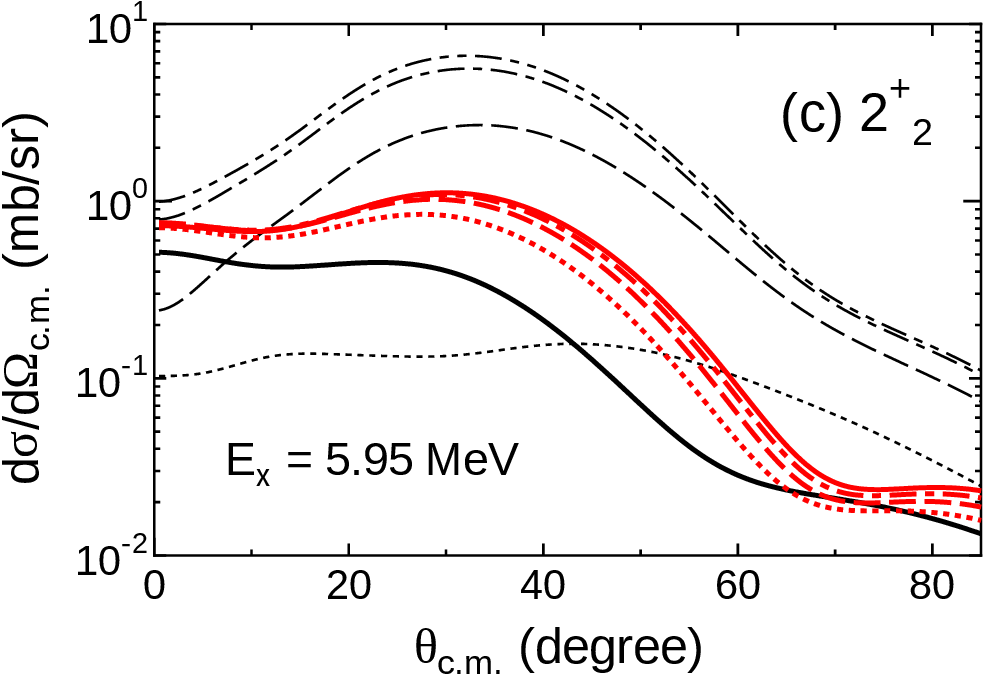}
\caption{\label{fig:59.4mev}
(a) Elastic cross section, (b) inelastic cross section for the 2$^+_1$ state, and (c) inelastic cross section for the 2$^+_2$ state of $^{10}$Be$+p$ system at $E/A =$ 59.4 MeV.
The meaning of the curves is the same as in Fig.~\ref{fig:trden}.
The experimental data are taken from \cite{EXFOR,cortina1997,iwasaki2000}.}
\end{figure}

\begin{figure}[h]
\centering
\includegraphics[width=8cm]{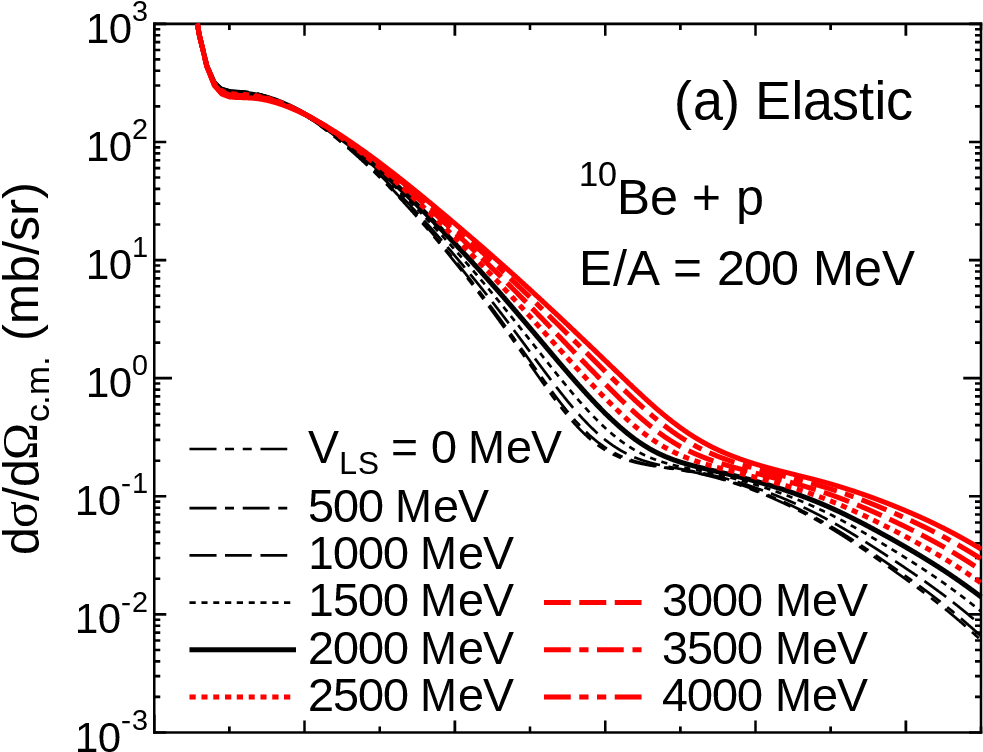}\\
\includegraphics[width=8cm]{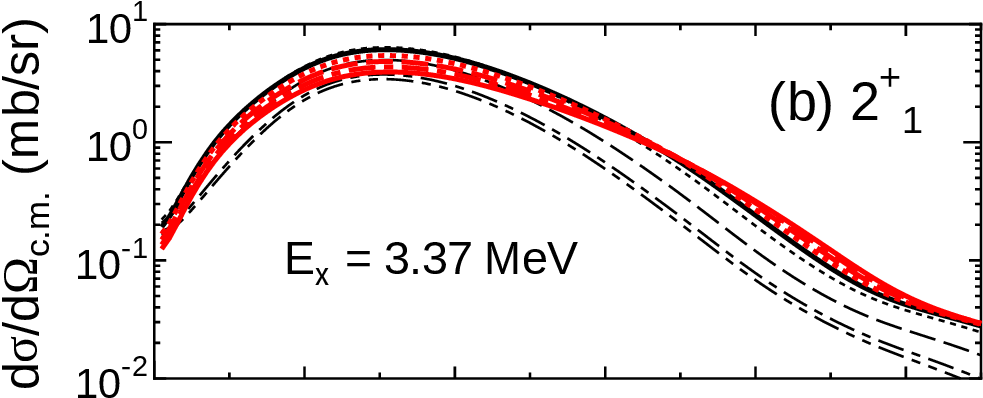}\\
\includegraphics[width=8cm]{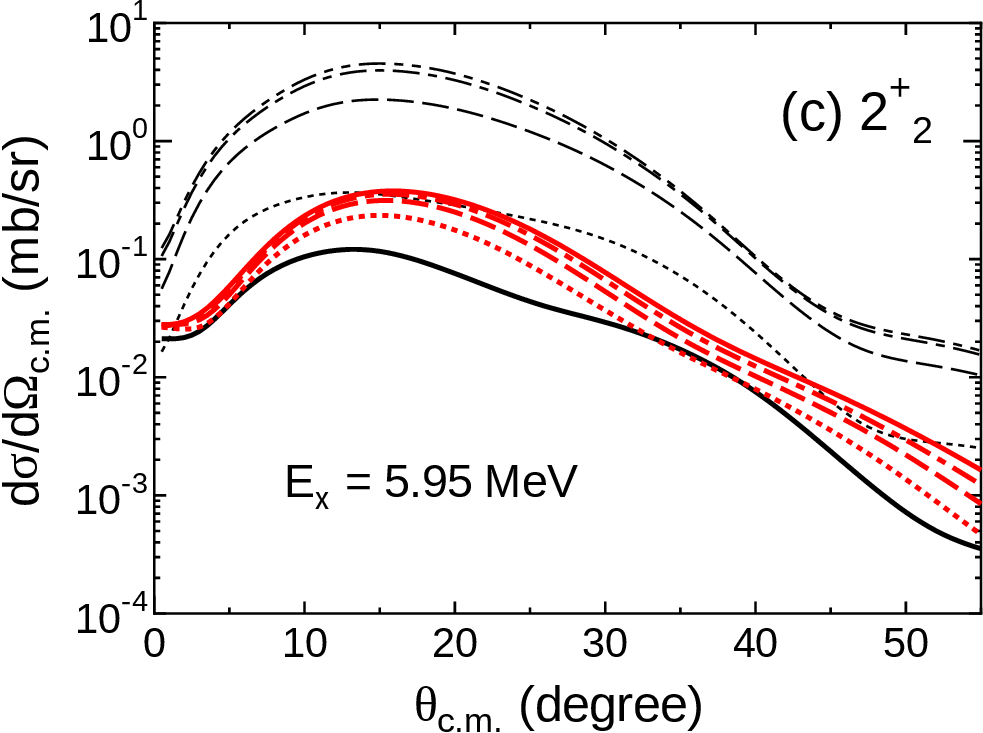}
\caption{\label{fig:200mev}
Same as Fig.~\ref{fig:59.4mev} but at $E/A =$ 200 MeV.}
\end{figure}

Figure~\ref{fig:59.4mev} shows the (a) elastic cross section, (b) inelastic cross section for the 2$^+_1$ state, and (c) inelastic cross section for the 2$^+_2$ state of the $^{10}$Be + $p$ system at $E/A =$ 59.4 MeV.
The experimental data are well reproduced in a wide range of $V_{\rm LS}$ values.
The change in the size of the ground state with varying $V_{\rm LS}$ slightly affects the elastic cross section.
The visible change in the transition strength and transition density from the 2$^+_1$ state to the 0$^+_1$ state results in a slight change in the inelastic cross section.
The sum of the proton and neutron parts of the transition density and $B$(IS2) (isoscalar component) is almost independent of $V_{\rm LS}$.
Therefore, the calculated inelastic cross sections are comparable to each other.
On the other hand, the calculated inelastic cross sections of the 2$^+_2$ state are greatly dependent on the $V_{\rm LS}$ value adopted in the structure calculation.
The drastic change in the inelastic cross section is caused by the difference in the transition strength and the form of the transition density, as shown in Figs.~\ref{fig:trst} and \ref{fig:trden}.
The behavior of the neutron part, in particular, is important.
Because of the sign reversal for the neutron part, the total transition density is drastically changed, giving rise to a drastic change in the inelastic cross section of the 2$^+_2$ state.
At $E/A =$ 59.4 MeV, the multistep effect (0$^+_1$ $\to$ 2$^+_1$ $\to$ 2$^+_2$) on the inelastic cross section of the 2$^+_2$ state cannot be ignored.
Therefore, we also calculated elastic and inelastic cross sections at high energy values at which the multistep effect is considered to be minor.
In fact, we confirmed that the multistep effect arising from the transition from the 2$^+_1$ state to the 2$^+_2$ state is minor at 200 MeV.
Figure~\ref{fig:200mev} shows the calculated (a) elastic cross section, (b) inelastic cross section for the 2$^+_1$ state, and (c) inelastic cross section for the 2$^+_2$ state at $E/A =$ 200 MeV.
The inelastic cross section for the 2$^+_2$ state is drastically influenced by the choice of the spin-orbit strength for the structure calculation, and this choice affects the inner structure of $^{10}$Be.
On the other hand, the elastic cross section and the inelastic cross section for the 2$^+_1$ state are almost constant.
This result implies that the degree of development of the dineutron correlation in $^{10}$Be can be observed for the 2$_2^+$ inelastic cross section.

\section{Summary and Outlook}
\label{sec:summary}

We combined microscopic structure and reaction calculations and discussed the changes in the cross section of the $^{10}$Be nucleus. For the structure calculations, we used a microscopic cluster model.
For the structure calculation, the inner structure of $^{10}$Be was artificially changed by changing the strength of the spin-orbit interaction $V_{\rm LS}$ = 0--4000 MeV.
During the artificial control, the 0$^+_1$ and 2$^+_1$ states exhibited similar behaviors in terms of the size and the expectation values of $<\bm{L} \cdot \bm{S}>$ and $<\bm{S}^2>$.
This result indicates that the 0$^+_1$ and 2$^+_1$ states have similar structures over all ranges of $V_{\rm LS}$.
However, the 2$^+_2$ state exhibited different behavior.
This difference is explained as follows:
The structure of the 2$^+_2$ state is similar to the structures of the 0$^+_1$ and 2$^+_1$ states, i.e., the 2$\alpha$+dineutron structure, when the spin-orbit interaction is weak.
However, the 2$^+_2$ state tends to have a different structure, i.e., 2$\alpha$+($\pi_{3/2}$)($\pi_{1/2}$), when the spin-orbit interaction is strong.

The strength of the spin-orbit interaction drastically affects the transition strength and the transition density from the excited state to the ground state.
The trends of the change of the transition density from the 2$^+_1$ or 2$^+_2$ states to the 0$^+_1$ state are completely different.
For the transition density from the 2$^+_1$ state to the 0$^+_1$ state, the behaviors of the proton and neutron parts are different.
On the other hand, the proton and neutron components of the transition density from the 2$^+_2$ state to the 0$^+_1$ state evolve from the negative sign to the positive sign together.
However, compared to the proton part, the neutron part shows an earlier sign reversal.

We applied the present wave functions to the elastic and inelastic scatterings by the proton target at $E/A =$ 59.4 and 200 MeV in the MCC calculation.
The change in the inner structure of the 0$^+_1$ and 2$^+_1$ states has a minor effect on the elastic cross section and the inelastic cross sections of the 2$^+_1$ state.
On the other hand, for the 2$^+_2$ state, the inelastic cross section underwent a drastic change.
Thus, we discovered the possibility of observing the degree of the development of the dineutron correlation in $^{10}$Be and its breaking for the 2$_2^+$ inelastic cross section.

In this study, we adopted the Volkov No.2 potential for the central part and the G3RS potential for the spin-orbit part to construct the $^{10}$Be nucleus with the 4-body $\alpha + \alpha + n + n$ cluster model.
However, we did not examine the dependence on the choice of the effective interaction, especially for the central part.
The Minnesota and Gogny D1S interactions were also applied to construct the $^{10}$Be nucleus~\cite{descouvemont2020,isaka2015}.
They reproduced the properties of the $^{10}$Be nucleus well.
Therefore, we consider that the drastic change in the inelastic cross section depends on the development of the dineutron correlation even if the effective interaction is replaced.
In future work, we will further develop our models and confirm that the
relation between the observable cross section and dineutron correlation
discussed herein holds in the cases of other effective interactions
for structure calculation.

In this study, we investigated the dineutron correlation and its relation to the spin-orbit potential.
We changed the strength of the spin-orbit interaction, which in turn changed of the total binding energy; in Ref.~\cite{itagaki2002}, the strength of the central potential was also adjusted to maintain a constant binding energy of the ground state.
Even after adjusting the binding energy, there was a drastic change in the transition from the 0$^+_1$ state to the 2$^+_2$ state.
In other words, the transition property between the 0$^+_1$ and 2$^+_2$ states is mainly governed by the spin-orbit interaction and is rather
independent of the total biding energy.
Therefore, we conclude that the development of the dineutron correlation in $^{10}$Be is sensitive to changes in the spin-orbit contribution, thereby resulting in drastic changes in the inelastic scattering of the 2$^+_2$ state.
We expect that more detailed analysis will confirm our results.

\vspace{2mm}
\acknowledgments
This work was supported by Japan Society for the Promotion of Science (JSPS) KAKENHI Grant Numbers JP20K03944 and JP21K03543.


\end{document}